\documentclass[a4paper]{article}
\author{Chiara Andr\`a\footnote{Dipartimento di Matematica, Universit\`a di Torino, Italy}, Luca Degiovanni\footnotemark[1], Guido Magnano\footnotemark[1]}
\date{}
\title{A trihamiltonian extension of the Toda lattice}

\usepackage{amssymb}
\usepackage{eufrak}
\usepackage{graphicx}

\newtheorem{proposition}{Proposition}

\newtheorem{theorem}[proposition]{Theorem}
\newtheorem{lemma}[proposition]{Lemma}

\newenvironment{Dim}%
{\par\medskip\upshape\noindent\textbf{Proof}:\small}%
{\hfill\normalsize{\textbf{c.v.d.}}\smallskip}

\newcommand{\Interi}{\mathbb{Z}}
\newcommand{\Reali}{\mathbb{R}}
\newcommand{\Gotico}[1]{\mathfrak{#1}}
\def\Uno{\mathrm{1\kern-2.4pt{I}}}

\newcommand{\BE}{\begin{equation}}
\newcommand{\EE}{\end{equation}}
\newcommand{\BAS}{\begin{eqnarray*}}
\newcommand{\EAS}{\end{eqnarray*}}

\newcommand{\D}{\mathrm{d}}
\def\aa{\mathtt{a}}
\newcommand{\Scal}[2]{(#1,#2)}
\newcommand{\SScal}[2]{(\!(#1,#2)\!)}
\newcommand{\Lie}[2]{[#1,#2]}
\newcommand{\LLie}[2]{[\![#1,#2]\!]}
\newcommand{\Pois}[2]{\{#1,#2\}}
\newcommand{\Adg}[1]{*#1}
\def\Op{\bullet}
\def\OOp{\odot}
\def\MR{$r$-matrix}
\def\Tr{\mathrm{tr}}
\def\TrKup{\mathfrak{Tr}}

\newcommand{\diff}[1]{{#1}^\prime}
\newcommand{\Tras}[1]{{#1}^\mathrm{t}}
\newcommand{\flm}{f_{\lambda\mu}}

\newcommand{\XSem}[1]{X_{#1}^{^{Sem}}}
\newcommand{\XSkl}[1]{X_{#1}^{^{Skl}}}

\def\PSem{P_{_{Sem}}}
\def\PSkl{P_{_{Skl}}}
\def\PLin{P_{_{Lin}}}
\newcommand{\ParSem}[2]{\{#1,#2\}_{_{Sem}}}
\newcommand{\ParSkl}[2]{\{#1,#2\}_{_{Skl}}}
\newcommand{\ParLin}[2]{\{#1,#2\}_{_{Lin}}}

\newcommand{\FrecciaAnd}[1]{
\put(0,0){\vector(3,2){15}}
\put(2,6){\tiny #1}
}
\newcommand{\FrecciaRit}[1]{
\put(15,0){\vector(-3,2){15}}
\put(8,6){\tiny #1}
}
\newcommand{\FrecciaGiu}[1]{
\put(7,19){\vector(0,-1){18}}
\put(8,8){\tiny #1}
}
\newcommand{\FrecciaP}{
\FrecciaAnd{$P$}
}
\newcommand{\FrecciaQ}{
\FrecciaRit{$Q$}
}
\newcommand{\FrecciaS}{
\FrecciaGiu{$S$}
}
\newcommand{\Scatola}[1]{
\put(0,0){\makebox(15,15){#1}}
}
\newcommand{\Lenard}[1]{
\put(0,0){\FrecciaP}
\put(30,0){\FrecciaQ}
\put(15,10){\Scatola{#1}}
}
\newcommand{\NucleoS}[1]{
\put(0,0){\FrecciaP}
\put(30,0){\FrecciaQ}
\put(15,25){\FrecciaS}
\put(15,10){\Scatola{#1}}
}
\newcommand{\BloccoS}[2]{
\put(0,35){\Scatola{#1}}
\put(0,15){\FrecciaS}
\put(0,0){\Scatola{#2}}
}
\newcommand{\BloccoQ}[2]{
\put(0,25){\Scatola{#2}}
\put(15,15){\FrecciaQ}
\put(30,0){\Scatola{#1}}
}
\newcommand{\BloccoP}[2]{
\put(0,0){\Scatola{#1}}
\put(15,15){\FrecciaP}
\put(30,25){\Scatola{#2}}
}
\newcommand{\MiniNucleoA}{
\put(0,0){\FrecciaP}
\put(15,0){\FrecciaQ}
\put(8,9){\FrecciaS}
}

\begin{document}

\maketitle
\abstract{\noindent A new Poisson structure on a subspace of the Kupershmidt algebra is defined. This Poisson structure, together with other two already known, allows to construct a trihamiltonian recurrence for an extension of the periodic Toda lattice with $n$ particles. Some explicit examples of the construction and of the first integrals found in this way are given.}

\section{Introduction}
The Toda lattice is a fundamental example of completely integrable system. It consists of $n$ 
particles on a line (in the non-periodic case) or on a circle (in the periodic
case), each interacting with its first neighbours through  an exponential repulsive force. The Hamiltonian of the non-periodic Toda lattice with $n$ particles thus is:
$$
H(q_1,..,q_n,p_1,..,p_n) = \frac{1}{2}\sum_{i=1}^{n}p_i^2 + \sum_{i=1}^{n-1} e^{q_i- q_{i+1}}.
$$
The periodic case is obtained allowing the second sum to run from $i=1$ to $n$ and setting $q_{n+1}=q_1$.

This lattice was first introduced by Toda in 1967; its integrability was demonstrated in 1974 by Flaschka, H\'enon and Manakov~\cite{Fla1974a,Hen1974,Man1974}. Moser~\cite{Mos1975b} found explicit
solution for the non-periodic case in 1975. In 1976 Bogoyavlensky~\cite{Bog1976} generalized the system by constructing analogous integrable lattices on any simple algebra. Kostant in~\cite{Kos1979b} showed an equivalence between
the integration of the Toda lattice and the representation theory of simple Lie algebras.  The Toda
lattice, either periodic or non-periodic, has been studied through \MR{} and bihamiltonian methods,
becoming a standard example in the theory of completely integrable systems. Many aspects of the
system were analyzed, in\-clu\-ding master symmetries, recursion operators, action-angle variables and
multiple Poisson brackets~\cite{Dam1994,DasO1989,DLNT1986}. Moreover many reformulations and
generalizations were given: beside the already cited constructions on simple algebras, it is worth
citing the relativistic Toda lattice~\cite{Rui1990}, the full Kostant--Toda lattice~\cite{EFS1991}, 
the extension of the system related to Lie algebroids~\cite{Meu2000,Meu2001} and the
Kupershmidt formulation of the periodic lattice in terms of shift operators~\cite{Kup1985}. In
recent years the problem of the separability of the Toda lattice has been approached with the
classical technique of point transformations~\cite{McLS2000} and from the point of view of
bihamiltonian theory~\cite{FMP2001}.

Both the periodic and the non-periodic Toda lattice admit a tri\-hamil\-to\-nian formulation (indeed a multi-hamiltonian one, see~\cite{Dam1994} and references therein), in the sense that three (compatible) Poisson structures
$P$,
$Q$, and
$S$ exist  such that, for a suitable set of Hamiltonians $\{h_i\}$, the recursion relations
\BE\label{tristandard}
X_i=P\D h_i = Q \D h_{i+1} = S\D h_{i-1}
\EE
hold. The Poisson tensors $P$, $Q$ and $S$ arise from a fairly general construction on Lie algebras,
equipped with a \MR{} and an underlying associative product~\cite{ORagn1989}. In particular for the
periodic Toda lattice this construction is performed on the Kupershmidt
algebra~\cite{MorPiz1996a,Kup1985}.

In the trihamiltonian recurrence (\ref{tristandard}), the Poisson tensor $S$ is in a sense redundant:
all the vector fields $X_i$ can be obtained using just the two tensors $P$ and $Q$. The
paper~\cite{triham} proposes a different kind of trihamiltonian recurrence: the Hamiltonians
aren't organized in a chain (labelled by only one index) but in a two-dimensional scheme (labelled
by two indices) and satisfy the relations
\BE\label{trinostro}
P\D h_{i,j} = Q \D h_{i+1,j}  = S\D h_{i,j+1}\,.
\EE
Clearly the two-dimensional scheme (\ref{trinostro}) can be thought as a set of one-dimensional
chains associated to the bihamiltonian pair $P$, $Q$ and linked together by the third Poisson
structure
$S$. In this framework the third structure is no more redundant, but allows to construct new Hamiltonians starting from a given one. Moreover it is possible to collect Hamiltonians belonging to different chains in a unique recursion scheme.
Some general properties of this kind of systems were investigated and a class of examples was
constructed in~\cite{triham} and~\cite{CalaGonone2004}, but until now very few systems were shown
to be trihamiltonian in this sense.\\

The aim of this article is to construct a new Poisson structure on a subspace of the Kupershmidt
algebra. This new Poisson tensor, together with the already known ones $P$ and $Q$, allows to construct a
trihamiltonian recurrence in the sense of (\ref{trinostro}) for an extension of the periodic Toda lattice.
The presented construction extends some partial results, obtained for the $3$-particles case in~\cite{tesiChiara,CalaGonone2004}, to the $n$-particles case.

\section{Background results}
\subsection{Poisson tensors and Hamiltonian vector fields}
Liouville's definition of complete integrability concerns Hamiltonian systems of mechanical origin, whereby the phase space is a cotangent bundle endowed with the canonical Poisson bracket. However, it has been thoroughly clarified several decades ago that relevant families of integrable hamiltonian systems are naturally defined in more general phase spaces, namely differentiable manifolds endowed with a (possibly de\-gen\-e\-rate) Poisson bracket, not derived from a natural symplectic structure but rather from other geometric or algebraic structures. A typical example is provided by the dual spaces of Lie algebras. We recall that a \emph{Poisson tensor} (or equivalently \emph{Poisson structure})
on a manifold $M$ is a bivector $P$ (i.e.~a skew-symmetric, contravariant, rank two tensor), such that the Schouten bracket $[P,P]$ vanishes. The latter condition ensures that the bracket defined by
$$
\{f,g\} = P(\D f,\D g)\, ,
$$
for an arbitrary pair of differentiable functions $(f, g)$ on $M$, fulfills the Jacobi identity and is therefore a Poisson bracket. A Poisson tensor $P$ associates to any function
$h$ a corresponding \emph{Hamiltonian vector field} $X_h$ given by the expression
$$
X_h = P\D h
$$
but, in contrast to the case of the canonical Poisson brackets of cotangent bundles, the vector field $X_h$ can be identically vanishing even if $h$ is nonconstant, because $P$ can be degenerate. A function $f$ such that $\D h\ne 0$ and $\D h\in\mathrm{Ker}\,P$ is called a \emph{Casimir function} for the Poisson structure $P$, and is in involution with any other function on $M$.

\subsection{Bihamiltonian structures}
As shown by Magri there is a systematic connection between the complete integrability of a Hamiltonian vector field and the existence of a second, independent Poisson structure, that is preserved by the same vector field. Two different Poisson tensors $P$ and $Q$ are said to be \emph{compatible} (in Magri's sense) if the linear combination (\emph{pencil})
$Q-\lambda P$ is a Poisson tensor for any $\lambda$. A particular case of Poisson pencil is obtained if the Lie derivative of $P$ along some vector field $X$ is also a Poisson tensor, because in that case the tensor $Q=\mathcal{L}_X P$ is easily proved to be compatible with $P$. Two compatible Poisson tensors define a
\emph{bihamiltonian structure}, and a vector field is called \emph{bihamiltonian} if it is
Hamiltonian with respect to both Poisson tensors: 
$$
X = P\D h = Q\D k\,.
$$
A \emph{Lenard chain} is a set of bihamiltonian vector fields $X_i = P\D h_i$ such that the
corresponding Hamiltonians satisfy the recursion relations
\BE\label{ricorrenza}
P\D h_i = Q\D h_{i+1}\,.
\EE
These recursion relations (\ref{ricorrenza}) can be represented by the diagram 
\BE\label{ric_biham}
\begin{picture}(250,45)
\put(30,0){\Scatola{$\D h_1$}} \put(45,15){\Lenard{$X_1$}}
\put(90,0){\Scatola{$\D h_2$}} \put(105,15){\Lenard{$\dots$}}
\put(150,0){\Scatola{$\D h_i$}} \put(165,15){\Lenard{$X_i$}}
\put(210,0){\Scatola{$\D h_{i+1}$}}
\end{picture}
\EE
In the sequel we shall omit the indication of the differential acting on the Hamiltonian functions, 
and simply draw $f {\buildrel P\over\rightarrow} X$ to mean $P:\D f \mapsto X$.
The recursion relations
(\ref{ricorrenza}) imply that all the vector fields belonging to a Lenard chain mutually commute
and hence their Hamiltonians are in involution with respect to both Poisson structures
\cite{M1997b}. Lenard recursion is thus the most effective way to produce integrable systems on a bihamiltonian manifold, and it turns out that both finite-dimensional and infinite-dimensional classical examples of integrable systems (including the case of solitonic equations) can be obtained, together with their symmetries and first integrals, as Lenard chains for suitable bihamiltonian structures. 

Whenever one of the Poisson tensors is degenerate constructing a Lenard chain by direct iteration becomes uneasy, but in that case one can look for a \emph{Casimir function of the Poisson pencil}, i.e.~a function $f_\lambda$, non-constant on $M$ and depending (as a formal power series) on the parameter $\lambda$, such that $(Q-\lambda P)\D f_\lambda\equiv 0$. It is straightforward to check that all the coefficients of the power expansion of $f_\lambda$ with respect to
$\lambda$ satisfy the recursion relations (\ref{ricorrenza}) and thus are in involution. If  two different Casimir functions of a Poisson pencil are present, the power expansion's coefficients of the first are not necessarily in involution with the ones of the second. But, if between these coefficients there is a Casimir function for one of the Poisson structures then, using the recurrence (\ref{ricorrenza}), the mutual involution with respect to both Poisson structures is again achieved.

\subsection{Trihamiltonian structures}
Although a pair of compatible Poisson structures provides a complete framework for the construction (or the investigation) of integrable systems, it has been recently suggested \cite{triham} that the existence of a third Poisson structure may be related to further properties, in turn connected to the \emph{algebraic integrability}. 
A \emph{trihamiltonian structure} on a manifold is given by three mutually compatible Poisson
structure $P$, $Q$ and $S$. A \emph{common Casimir function} for a trihamiltonian structure is a
function $\flm$ (parametrically dependent on two parameters) such that
\BE\label{Cas_Com}
\left\{
\begin{array}{ccc}
(Q -\lambda P)\D\flm &=& 0 \\
(S -\mu P)\D\flm &=& 0
\end{array}
\right.
\EE
As a matter of fact, a generic trihamiltonian structure may not admit at all such a function (see \cite{triham} for a simple example). Although a precise classification of trihamiltonian structures admitting common Casimir functions is an open problem at present, the main point of introducing a third Poisson structure is exactly the possible existence of nontrivial (and complete, in a suitable sense) solutions to (\ref{Cas_Com}). If the
function $\flm$ exists, the coefficients of its powers expansion
$\flm=\sum f_{ij}\lambda^i\mu^j$ obey the recursion rules
\begin{equation}\label{rec_triham}
X_{ij}=P\D f_{ij} =  Q\D f_{i+1,j} = S\D f_{i,j+1}\,,
\end{equation}
which can be represented by the following recursion diagram:
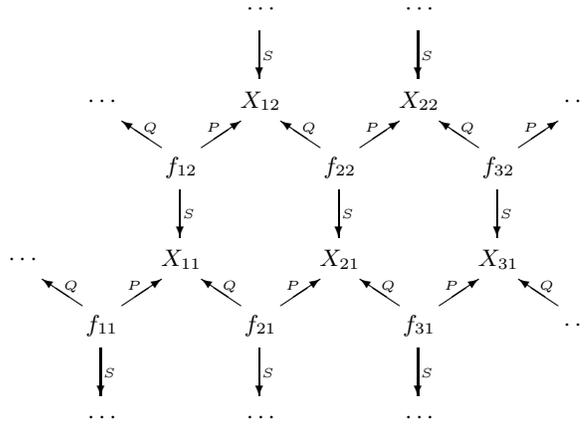
\begin{center}
\begin{picture}(240,165)
\put(30,0){\BloccoS{}{$\cdots$}}
\put(90,0){\BloccoS{$f_{21}$}{$\cdots$}}
\put(150,0){\BloccoS{$f_{31}$}{$\cdots$}}
\put(0,35){\BloccoQ{$f_{11}$}{$\cdots$}} \put(45,50){\NucleoS{$X_{11}$}}
\put(105,50){\NucleoS{$X_{21}$}} \put(165,50){\NucleoS{$X_{31}$}} \put(210,35){\Scatola{$\cdots$}}
\put(30,95){\BloccoQ{$f_{12}$}{$\cdots$}} \put(75,110){\NucleoS{$X_{12}$}}\put(120,95){\Scatola{$f_{22}$}}\put(135,110){\NucleoS{$X_{22}$}}
\put(180,95){\BloccoP{$f_{32}$}{$\cdots$}}
\put(90,155){\Scatola{$\cdots$}}\put(150,155){\Scatola{$\cdots$}}
\end{picture}
\end{center}

Similarly to the bihamiltonian case, the recursion relations (\ref{rec_triham}) imply that the coefficients $f_{ij}$ are mutually in involution with respect to each of the three Poisson structures~\cite{triham}. The vector fields $X_{ij} = P\D f_{ij} =  Q\D f_{i+1,j} = S\D f_{i,j+1}$ are trihamiltonian vector fields. 

It is worth observing that the power expansion of a common Casimir function contain only non-negative powers (i.e. it is a formal Taylor series in both $\lambda$ and $\mu$) only if $Q$ and $S$ are degenerate Poisson tensors:  for every power of $\lambda$ the least order coefficient in $\mu$ must be a Casimir function for $S$, and the least order coefficient in $\lambda$ must be a Casimir function for $Q$ for every power of $\mu$.

Moreover the apparently different role played by $P$ in equations (\ref{Cas_Com}) depends only on the choice of the parameters $\lambda$ and $\mu$. In fact, setting $\lambda=\nu_P/\nu_Q$ and $\mu=\nu_P/\nu_S$ in (\ref{Cas_Com}), it is clear that every choice of two of the following three equations
$$
\left\{
\begin{array}{ccc}
(\nu_Q Q -\nu_P P)\D f &=& 0 \\
(\nu_S S -\nu_P P)\D f &=& 0 \\
(\nu_Q Q -\nu_S S)\D f &=& 0
\end{array}
\right.
$$
implies the remaining one.

A remarkable feature of trihamiltonian systems is that, if a suitable set of  vectors fields $Z_\alpha$ is found, the common Casimir function allows to calculate a set of separation coordinates for all its coefficients. The technical conditions under which this result holds are stated in the following proposition (see \cite{triham} for a proof):
\begin{proposition}
Let  $P$, $Q$ and $S$ three mutually compatible Poisson structure of constant corank $k$ on a manifold of dimension $2n+k$, let $\flm$ a polynomial common Casimir function of the two pencils $Q-\lambda P$ and $S-\mu P$ containing $k$ Casimir function $c^\alpha$ of $P$ and let $h^\alpha$, $k^\alpha$ the coefficients of $\flm$ such that $P\D h^\alpha=Q c^\alpha$ and $P\D k^\alpha=S c^\alpha$ respectively. Given $k$ independent vector field $Z_\alpha$ let the following conditions holds:
\begin{itemize}
\item $Z_\alpha(c^\beta)=\delta^\beta_\alpha$ and $\Lie{Z_\alpha}{Z_\beta}=0$;
\item if $Q_d=Q+L_{h^\alpha Z_\alpha}P$ and $S_d=S+L_{k^\alpha Z_\alpha}P$ then $L_{Z_\alpha}P=L_{Z_\alpha}Q_d=L_{Z_\alpha}S_d=0$;
\item $Z_\alpha(Z_\beta(\flm))=0$;
\item the polynomials $\sigma_\alpha=Z_\alpha(\flm)$ have $2n$ functionally independent common roots $\{\lambda_i,\mu_i\}$;
\item the equality
$$
{\Pois{\sigma_\alpha}{\sigma_\beta}}_P\Big|_{\lambda_i,\mu_i}=\left.\left(
\frac{\partial \sigma_\alpha}{\partial\lambda}\frac{\partial \sigma_\beta}{\partial\mu}-
\frac{\partial \sigma_\beta}{\partial\lambda}\frac{\partial \sigma_\alpha}{\partial\mu}
\right)\right|_{\lambda_i,\mu_i}
$$
holds and $\forall i \;\exists \mbox{ a pair } (\alpha,\beta)$ for which both sides do not vanish identically.
\end{itemize}
Then, on any symplectic leaf of $P$, the $2n$ functions $\{\lambda_i,\mu_i\}$ are canonical coordinates and exists $k$ (not necessarily distinct) polynomials $p_i(\lambda,\mu)$, with constant coefficients, such that
$$
f_{\lambda_i\mu_i}=p(\lambda_i,\mu_i)\,.
$$
Hence the $k$ functions $W_i=\flm-p_i(\lambda,\mu)$ establish the separation in Sklyanin sense of all the coefficients of $\flm$ in the coordinates $\{\lambda_i,\mu_i\}$.
\end{proposition}

\subsection{Hamiltonian systems on Lie algebras}\label{LieAlgebre}
A particularly important example of Poisson manifold is any Lie algebra $\Gotico{g}$ equipped with a scalar product $\Scal{\cdot}{\cdot}$. In this case, it is well known that given the functions $f,g$ on $\Gotico{g}$ and their gradient $\nabla f, \nabla g$ with respect to $\Scal{\cdot}{\cdot}$, the bracket
$$
\{f,g\}(L)=\Scal{L}{\Lie{\nabla f}{\nabla g}}
$$
is a Poisson bracket called the \emph{Lie--Poisson bracket}. An analogous Poisson bracket is defined on the dual $\Gotico{g}^\ast$ of the Lie algebra. In this paper, the scalar product on $\Gotico{g}$ is supposed to be \emph{invariant}, i.e.~for every $A,B,C\in\Gotico{g}$ to satisfy
$$
\Scal{A}{\Lie{B}{C}}=\Scal{B}{\Lie{C}{A}}\,.
$$

If the Lie bracket $\Lie{\cdot}{\cdot}$ is the commutator obtained from an underlying associative algebra structure and if a \MR{} is present, i.e. a linear map $R:\Gotico{g}\to\Gotico{g}$ satisfying the Yang--Baxter equation
\BE\label{Yang-Baxter}
[R(A),R(B)]-R([R(A),B]+[A,R(B)]) = - [A,B]
\EE
then a bihamiltonian structure can be defined. Indeed, if $\Gotico{g}$ is an associative algebra, then the following two natural operations are defined:
\begin{eqnarray}
[A,B]&=&AB-BA \label{commLie}\,;\\
A\Op B &=& \frac{AB+BA}{2}\,.\label{prodJordan}
\end{eqnarray}
It is worth observing that the product $\Op$ is commutative, but in general not associative and satisfies the relation
\BE\label{der_prop}
\Lie{A}{B\Op C}=\Lie{A}{B}\Op C +B\Op\Lie{A}{C}\,.
\EE
This richer structure allow, as proved in~\cite{ORagn1989}, to state a sufficient condition for the existence of a bihamiltonian structure on $\Gotico{g}$:
\begin{proposition}\label{prop_ORag}
If a linear map $R:\Gotico{g}\to\Gotico{g}$ exists, such that both $R$ and $R_a=(R-\Tras{R})/2$ (where $\Scal{A}{R(B)}=\Scal{\Tras{R}(A)}{B}$) satisfy the Yang--Baxter equation (\ref{Yang-Baxter}), then the two brackets
\begin{eqnarray}
\ParSem{f}{g}&=&(L,[R(\nabla f),\nabla g]+[\nabla f,R(\nabla g)]) \label{Sem1}\\
\ParSkl{f}{g}&=&(L,[R(L\Op\nabla f),\nabla g]+[\nabla f,R(L\Op\nabla g)]) \label{Skl1}
\end{eqnarray}
are compatible Poisson brackets on $\Gotico{g}$.
\end{proposition}
The linear Poisson structure (\ref{Sem1}) and the quadratic one (\ref{Skl1}) will be called respectively \emph{Sem\"enov-Tian-Shansky bracket} and \emph{Sklyanin bracket}.
A standard way to produce a \MR{} on a Lie algebra $\Gotico{g}$ is to decompose it into two Lie subalgebras $\Gotico{g}_+$ and $\Gotico{g}_-$. In this case (known as the \emph{split case}~\cite{ReySem1994}) the map
$$
R=\frac{1}{2}(\Pi_+-\Pi_-)
$$
is a \MR{}. Defining through the invariant scalar product, the orthogonal decomposition $\Gotico{g}=\Gotico{g}^+\oplus\Gotico{g}^-$, with $\Gotico{g}^\pm=(\Gotico{g}_\pm)^\perp$, the Hamiltonian vector fields generated by the two structures (\ref{Sem1}) and (\ref{Skl1}) can be set in both the two forms
\begin{eqnarray}\label{CampiSem}
\XSem{f} & = & [L, (\nabla f)_\pm]-[L,\nabla f]^\pm \\
\label{CampiSkl}
\XSkl{f} & = & [L,(L\Op\nabla f)_\pm]-L\Op[L,\nabla f]^\pm\,.
\end{eqnarray}

This  introductory section is concluded with a remark on the Lie derivative of a Lie--Poisson bracket:
\begin{lemma}\label{derivo}
Let $X$ be a vector field on a Lie algebra $\Gotico{g}$ and $P$ the Poisson tensor associated to the Lie--Poisson bracket. Then the Poisson bracket obtained
from $\mathcal{L}_X P$ is
$$
{\{f,g\}}_{\mathcal{L}_X P}=\Scal{X}{\Lie{F}{G}}-\Scal{L}{\Lie{\Tras{\diff{X}}\cdot
F}{G}+\Lie{F}{\Tras{\diff{X}}\cdot G}}\,,
$$
where ${\diff{X}}\cdot M$ is the derivative of $X$ (thought as a function from the vector space $\Gotico{g}$ to itself) in the direction of the vector $M$, and $F=\nabla f$, $G=\nabla g$.
\begin{Dim}
The expression of the Poisson bracket associated to the Lie derivative of a Poisson tensor is:
$$
{\{f,g\}}_{\mathcal{L}_X P}=X({\{f,g\}}_P)-{\{X(f),g\}}_P-{\{f,X(g)\}}_P\,.
$$
Considering the Lie--Poisson structure one has
\BAS
X({\{f,g\}}_P)(L) &=& \left.\frac{\D}{\D \tau}\right|_{\tau=0}\!\!\!\Scal{L+\tau X}{\Lie{F(L+\tau X)}{G(L+\tau X)}} \\
&=& \Scal{X}{\Lie{F}{G}}+\Scal{L}{\Lie{\diff{F}\cdot X}{G}}+\Scal{L}{\Lie{F}{\diff{G}\cdot X}}
\EAS
and, because $X(f)=\Scal{X}{F}$,
\BAS
\Scal{\nabla X(f)}{M} &=& \left.\frac{\D}{\D \tau}\right|_{\tau=0}\!\!\!\Scal{X(L+\tau M)}{F(L+\tau M)} \\
&=& \Scal{X}{\diff{F}\cdot M}+\Scal{\diff{X}\cdot M}{F} \\
&=& \Scal{M}{\Tras{\diff{F}}\cdot X+\Tras{\diff{X}}\cdot F}
\EAS
thus $\nabla X(f)=\Tras{\diff{F}}\cdot X+\Tras{\diff{X}}\cdot F$.
Because $F=\nabla f$, one obtains
\BAS
\Scal{\Tras{\diff{F}}\cdot M}{N} &=& \Scal{M}{\diff{F}\cdot N} \\
&=& \left.\frac{\D}{\D \tau}\right|_{\tau=0}\!\!\!\Scal{M}{F(L+\tau N)}\\
&=&\left.\frac{\D}{\D \tau}\right|_{\tau=0}\left.\frac{\D}{\D \sigma}\right|_{\sigma=0}\!\!\!f(L+\tau N+\sigma M)\\
&=&\left.\frac{\D}{\D \sigma}\right|_{\sigma=0}\!\!\!\Scal{N}{F(L+\sigma M)}\\
&=& \Scal{N}{\diff{F}\cdot M}
\EAS
thus the relation $\Tras{\diff{F}}\cdot M = \diff{F}\cdot M$ holds. On the other hand, note that
$\Tras{\diff{X}}\cdot M$ is not generally equal to $\diff{X}\cdot M$. The
thesis is obtained with a simple substitution.
\end{Dim}
\end{lemma}

\section{A new trihamiltonian structure}

\subsection{The Kupershmidt algebra}
The Kupershmidt algebra $\mathcal{K}_n$ is an associative algebra with unity introduced in~\cite{Kup1985} in order to establish an algebraic framework for the periodic Toda lattice. The elements of $\mathcal{K}_n$ are formal Laurent series in $\Delta$ with coefficients in the ring of $n$-periodic sequences $\sigma=(\sigma|_1,\ldots, \sigma|_n)$, with componentwise operations. The multiplication between powers of $\Delta$ is commutative and given by $\Delta^l\Delta^k=\Delta^{l+k}$, the identity is $\Delta^0$ and the multiplication between powers of $\Delta$ and sequences is determined by the formula
$$
\Delta^k\sigma=\sigma^{[k]}\Delta^k\,,
$$
where the sequence $\sigma^{[k]}$ is given by 
$$
\sigma^{[k]}=(\sigma|_{k+1},\ldots,\sigma|_n,\sigma|_1,\ldots,\sigma|_k)\,.
$$
Because of to this relation $\Delta$ can be interpreted as a shift operator on the sequences of period $n$.

\begin{figure}
\begin{eqnarray*}
\sigma & \longrightarrow & \left(
\begin{array}{cccc}
\sigma_1 & 0 & \cdots & 0\\
0 & \sigma_2 & \ddots & \vdots\\
\vdots & \ddots & \ddots & 0\\
0 & \cdots & 0 & \sigma_n
\end{array} \right) \\
\sigma \Delta & \longrightarrow & \left(
\begin{array}{cccc}
0 & \sigma_1 &\cdots & 0\\
0 & 0 & \ddots & \vdots\\
\vdots & \ddots & \ddots & \sigma_{n-1}\\
\lambda \sigma_n & 0 & \cdots & 0
\end{array} \right) \\
%
%
& \vdots \\
\sigma \Delta^{n-1} & \longrightarrow & \left(
\begin{array}{cccc}
0 & \cdots & 0 & \sigma_1\\
\lambda \sigma_2 & 0 & \cdots & 0\\
\vdots & \ddots & \ddots & \vdots\\
0 & \cdots & \lambda \sigma_n & 0\\
\end{array} \right) \\
\sigma \Delta^n & \longrightarrow & \left(
\begin{array}{cccc}
\lambda \sigma_1 & 0 & \cdots & 0 \\
0 & \lambda \sigma_2 & \ddots & \vdots \\
\vdots & \ddots & \ddots & 0 \\
0 & \cdots & 0 & \lambda \sigma_n
\end{array} \right)\\
\Delta^{-1} \sigma & \longrightarrow & \left(
\begin{array}{cccc}
0 & \cdots & 0 & \lambda^{-1} \sigma_n \\
\sigma_1 & 0 & \cdots & 0\\
\vdots & \ddots & \ddots & \vdots\\
0 & \cdots & \sigma_{n-1} & 0
\end{array} \right) \\
%
%
& \vdots \\
\Delta^{-n+1} \sigma & \longrightarrow & \left(
\begin{array}{cccc}
0 & \lambda^{-1} \sigma_2 & \cdots & 0\\
\vdots & \ddots & \ddots & \vdots\\
0 & & \ddots & \lambda^{-1} \sigma_n \\
\sigma_1 & 0 & \cdots & 0
\end{array} \right) \\
\Delta^{-n} \sigma & \longrightarrow & \left(
\begin{array}{cccc}
\lambda^{-1} \sigma_1 & 0 & \cdots & 0 \\
0 & \lambda^{-1} \sigma_2 & \ddots & \vdots \\
\vdots & \ddots & \ddots & \vdots \\
0 & \cdots & 0 & \lambda^{-1} \sigma_n
\end{array} \right)
\end{eqnarray*}
\caption{Isomorphism between the algebras $\mathcal{K}(n)$ and $\Gotico{gl}(n)(\lambda,\lambda^{-1}]$}
\label{isoKup}
\end{figure}

In~\cite{MorPiz1996a} the algebra $\mathcal{K}_n$ was shown to be isomorphic to the loop algebra $\Gotico{gl}(n)(\lambda,\lambda^{-1}]$ of formal Laurent series with matrix coefficients, through the map shown in Figure~\ref{isoKup}. On the algebra $\mathcal{K}_n$ one can define a linear involution $\ast$ by $\ast(\sigma\Delta^k)=\Delta^{-k}\sigma$ and the trace
$$
\TrKup\left(\sum_{k\in\Interi} \sigma_k\Delta^k\right)=\sum_{i=1}^n \sigma_0|_i\,;
$$
under the isomorphism between $\mathcal{K}_n$ and $\Gotico{gl}(n)(\lambda,\lambda^{-1}]$ they are mapped in the involution $\ast(L_\lambda)=(L_{\lambda^{-1}})^t$ and in the trace
$$
\TrKup\left(\sum_{k\in\Interi}L_k\lambda^k\right)=\Tr(L_0)\,.
$$
From now on the two algebras $\mathcal{K}_n$ and $\Gotico{gl}(n)(\lambda,\lambda^{-1}]$ are identified and indicated with $\Gotico{G}$. The choice of the representation used will be left to the context.

On the associative algebra $\Gotico{G}$ one can define the commutator (\ref{commLie}), the commutative product (\ref{prodJordan}) and the invariant scalar product
\BE\label{prodinv}
\Scal{L}{M}=\TrKup(LM)
\EE
together with its restriction (indicated with the same symbol) to the na\-tu\-ral immersion of $\Gotico{gl}(n)$ in $\Gotico{gl}(n)(\lambda,\lambda^{-1}]$. Moreover, it is possible to find a decomposition of $\Gotico{G}$ into two Lie subalgebras and thus to find a \MR{}: following~\cite{MorPiz1996a} one introduce the subspaces
\begin{itemize}
\item $\Gotico{G}_+$: skewsymmetric elements, $\ast L=-L$;
\item $\Gotico{G}^+$: symmetric elements, $\ast L=L$;
\item $\Gotico{G}_k$: elements of degree $k$, with powers of $\Delta$ not greater than $k$; 
\item $\Gotico{G}_-=\Gotico{G}_0$; 
\item $\Gotico{G}^-=\Gotico{G}_{-1}$.
\end{itemize}
As easily seen, $\Gotico{G}_\pm$ are two subalgebras of $\Gotico{G}$ and $\Gotico{G}=\Gotico{G}_+\oplus\Gotico{G}_-=\Gotico{G}^+\oplus\Gotico{G}^-$, thus the linear map $R=\Pi_+-\Pi_-$ is a \MR{}. It can be proved~\cite{MorPiz1996a} that its skewsymmetric part $(R-\Tras{R})/2$ also satisfies the Yang--Baxter equation. Proposition~\ref{prop_ORag} implies that the two brackets (\ref{Sem1}) and (\ref{Skl1}) endow $\Gotico{G}$ with a bihamiltonian structure. Lastly, the relation $\Gotico{G}^\pm=(\Gotico{G}_\pm)^\perp$ holds, thus the Hamiltonian vector fields can be set in the form (\ref{CampiSem}) and (\ref{CampiSkl}).

A Poisson structure is said to be \emph{reducible by restriction} on a given submanifold if any Hamiltonian vector fields is tangent to it, when evaluated on the submanifold. In other words, if the submanifold is invariant with respect to any Hamiltonian vector field. On the algebra $\Gotico{G}$ a reduction procedure for the Poisson structures (\ref{Sem1}) and (\ref{Skl1}) can be performed: the bihamiltonian structure of  $\Gotico{G}$ is reducible on $\Gotico{G}_k^+=\Gotico{G}^+\cap\Gotico{G}_k$ for any $k$. This is the content of the following theorem, explicitly proved in~\cite{MorPiz1996a} for the case $\Gotico{G}_1^+$ and somehow already present in~\cite{ORagn1989}.
\begin{theorem}
Given any function $f:\Gotico{G}\to\Reali$, both the subspaces $\Gotico{G}^+$ and $\Gotico{G}_k$ are invariant with respect to the corresponding Hamiltonian vector fields (\ref{CampiSem}) and (\ref{CampiSkl}). As a consequence, the two Poisson structures (\ref{Sem1}) and (\ref{Skl1}) are reducible by restriction on $\Gotico{G}^+$, $\Gotico{G}_k$ and on their intersection $\Gotico{G}_k^+$.
\begin{Dim}
Firstly, the reducibility of the Sem\"enov-Tian-Shansky structure is proved. Set $F=\nabla f$, the invariance of $\Scal{\cdot}{\cdot}$ implies
$$
\Scal{M}{\Lie{L}{F_+}} =  \Scal{L}{\Lie{F_+}{M}} 
$$
and being $\Gotico{G}_+$ a subalgebra orthogonal to $\Gotico{G}^+$, if $M\in\Gotico{G}_+$ and $L\in\Gotico{G}^+$ then
$$
\Scal{M}{\Lie{L}{F_+}} = 0 \quad \forall M\in\Gotico{G}_+ \quad\Rightarrow\quad \Lie{L}{F_+}\in\Gotico{G}^+\,.
$$
From the expression (\ref{CampiSem}) one gets that $\XSem{f}\in\Gotico{G}^+$. On the other hand, if $L\in\Gotico{G}_k$ then $[L,F_-]$ has maximum degree $k$, because $F_-\in\Gotico{G}_0$, and $[L,F]^-\in\Gotico{G}_{-1}$. From (\ref{CampiSem}) one gets that $\XSem{f}\in\Gotico{G}_k$.

Analogously, from (\ref{CampiSkl}) one can prove the reducibility of the Sklyanin Poisson structure. Observing that, if $L\in\Gotico{G}^+$, then $[L,(L\Op\nabla f)_+]\in\Gotico{G}^+$, and that $\Gotico{G}^+\Op\Gotico{G}^+\subset\Gotico{G}^+$ one obtains that $\XSkl{f}\in\Gotico{G}^+$. Moreover, because $L\in\Gotico{G}_k$ either $[L,(L\Op\nabla f)_-]$ or $L\Op[L,\nabla f]^-$ have maximum degree $k$.

The reducibility of the two Poisson structures on the intersection $\Gotico{G}^+_k=\Gotico{G}^+\cap\Gotico{G}_k$ follows straightforwardly.
\end{Dim}
\end{theorem}

\subsection{The third Hamiltonian structure}
The trihamiltonian structure proposed in this paper is defined on the subspace $\Gotico{G}_{n-1}^+$ of the algebra $\Gotico{G}$. An element of $\Gotico{G}_{n-1}^+$ can be expressed in one of the two forms:
\BAS
&&\Delta^{n-1}\sigma_{n-1}+\cdots+\Delta\sigma_1+\sigma_0+\sigma_1\Delta+\cdots+\sigma_{n-1}\Delta^{n-1}\\
&& L_0 +L_1\lambda+L_1^t\lambda^{-1}
\EAS
where $L_0$ is a symmetric matrix and $L_1$ is a strictly lower triangular matrix (both $L_0$ and $L_1$ do not depend on $\lambda$).

A very natural isomorphism $\Phi$ between the vector space
$\Gotico{G}_{n-1}^+$ and the vector space $\Gotico{gl}(n)$ exists: it can be constructed using the decomposition
$$
L_\lambda=L_0+L_1\lambda+L_1^t\lambda^{-1}
$$
for the elements of $\Gotico{G}_{n-1}^+$. The map $\Phi$ is thus given by
\BE\label{isomorfismo}
\begin{array}{rcl}
\Phi : \Gotico{G}_{n-1}^+ &\to& \Gotico{gl}(n) \\
L_\lambda &\mapsto& L=L_0+L_1\!-\!L_1^t
\end{array}
\EE
and it is clearly invertible. The symmetric and skewsymmetric part
of $L=\Phi(L_\lambda)$ are indicated with $L_S=(L+L^t)/2$ and $L_A=(L-L^t)/2$, respectively. It is important to observe that $\Phi$ is just an isomorphism between vector spaces because, although $\Gotico{gl}(n)$ has a structure of associative algebra, $\Gotico{G}_{n-1}^+$ lacks this structure.

The map $\Phi$ induces a scalar product on $\Gotico{gl}(n)$:
\begin{lemma}
Let $\Scal{\cdot}{\cdot}$ be the scalar product~(\ref{prodinv}) on $\Gotico{G}$, then the relation
$$
\Scal{L_\lambda}{M_\lambda}=\SScal{L}{M}
$$
holds for all $L_\lambda,\, M_\lambda\in\Gotico{G}_{n-1}^+$, where $L=\Phi(L_\lambda)$, $M=\Phi(M_\lambda)$ and
$$
\SScal{L}{M}=\Scal{L_S}{M_S}-\Scal{L_A}{M_A}=\Tr(LM^t)\,.
$$
\begin{Dim}
Using the decomposition $L_\lambda=L_0+L_1\lambda+L_1^t\lambda^{-1}$ and the scalar product~(\ref{prodinv}) one obtains
\BAS
\Scal{L_\lambda}{M_\lambda} &=& \Tr(L_0M_0+L_1M_1^t+L_1^tM_1) \\
&=& \Tr(L_0M_0+L_1M_1^t-L_1M_1+L_1^tM_1-L_1^tM_1^t) \\
&=& \Tr(L_SM_S)-\Tr(L_AM_A)\,.
\EAS
The last expression for the scalar product on $\Gotico{gl}(n)$ follows straightforwardly from the
definitions of $L_S$ and $L_A$.
\end{Dim}
\end{lemma}

A non-standard Lie bracket, fundamental for the construction of the new Poisson bracket, can be defined on the space $\Gotico{gl}(n)$:
\begin{lemma}\label{Lie doppia}
The bracket on $\Gotico{gl}(n)$
$$
\LLie{L}{M}=\Lie{L_S}{M_S}-\Lie{L_A}{M_A}-\Lie{L_S}{M_A}-\Lie{L_A}{M_S}
$$
is a Lie bracket and the relation
$$
\SScal{L}{\LLie{M}{N}}=\SScal{M}{\LLie{N}{L}}
$$
holds for every $L,M,N\in\Gotico{gl}(n)$.
\begin{Dim}
First it is proved that $\LLie{\cdot}{\cdot}$ is a Lie bracket: just the Jacobi identity is checked because skewsymmetry and bilinearity are trivially satisfied.
\BAS
\LLie{L}{\LLie{M}{N}}\!\!\!\!&=&\!\!\!\!\Lie{L_A}{\Lie{M_A}{N_S}\!+\!\Lie{M_S}{N_A}}  + \Lie{L_S}{\Lie{M_A}{N_A}\!-\!\Lie{M_S}{N_S}}+\\
&&\!\!\!\!\Lie{L_A}{\Lie{M_A}{N_A}\!-\!\Lie{M_S}{N_S}}  -\Lie{L_S}{\Lie{M_A}{N_S}\!+\!\Lie{M_S}{N_A}}\,,
\EAS
cyclically permuting $L$, $M$ and $N$ and using the Jacobi identity for the ordinary commutator, the thesis is proved.
On the other hand, using the invariance of $\Scal{\cdot}{\cdot}$ with respect to the ordinary commutator one gets
\BAS
-\SScal{L}{\LLie{M}{N}}\!\!\!\!&=&\!\!\!\!\Scal{L_S}{\Lie{M_S}{N_A}\!+\!\Lie{M_A}{N_S}} \!+\! \Scal{L_A}{\Lie{M_S}{N_S}\!-\!\Lie{M_A}{N_A}} \\
&=&\!\!\!\!\Scal{M_S}{\Lie{N_A}{L_S}\!+\!\Lie{N_S}{L_A}}+\Scal{M_A}{\Lie{N_S}{L_S}\!-\!\Lie{N_A}{L_A}} \\
&=&\!\!\!\!-\SScal{M}{\LLie{N}{L}}\,.
\EAS
\end{Dim}
\end{lemma}

It is very useful, in order to make the notation more compact, to introduce the commutative product
\BE\label{prod_doppio}
L\OOp M = L_S \Op M_S - L_A \Op M_A + L_A\Op M_S + L_S \Op M_A\,.
\EE
From the definition of $\OOp$ it follows that, if $M$ is symmetric, then $L\OOp M= L\Op M$. Thus, in particular the matrix $\Uno$ is the identity also for this product.
\begin{lemma}\label{lemmamio1}
The commutative product (\ref{prod_doppio}) satisfies the properties
\BAS
&&\SScal{L\OOp M}{N}=\SScal{L}{M\OOp N}\\
&&\LLie{L}{M\OOp N}=\LLie{L}{M}\OOp N + M\OOp\LLie{L}{N}\,.
\EAS
\begin{Dim}
The first property is easily proved:
\BAS
\SScal{L\OOp M}{N}\!\!\!\!&=&\!\!\!\!\Scal{L_S\Op M_S -L_A \Op M_A}{N_S} - \Scal{L_A\Op M_S+L_S\Op M_A}{N_A} \\
&=&\!\!\!\!\Scal{L_S}{M_S \Op N_S -M_A \Op N_A} - \Scal{L_A}{M_A \Op N_S + M_S \Op N_A} \\
&=&\!\!\!\!\SScal{L}{M \OOp N}\,.
\EAS
The second property requires a bit more work:
\BAS
\LLie{L}{M\OOp N}\!\!\!\!&=&\!\!\!\!
\Lie{L_S}{M_S\Op N_S-M_A\Op N_A}-\Lie{L_A}{M_S\Op N_A+M_A\Op N_S}-\\
&&\!\!\!\!\Lie{L_S}{M_S\Op N_A+M_A\Op N_S}-\Lie{L_A}{M_S\Op N_S-M_A\Op N_A}\\
&=&\!\!\!\!
(\Lie{L_S}{M_S}-\Lie{L_A}{M_A}-\Lie{L_S}{M_A}-\Lie{L_A}{M_S})\Op N_S \\
&&\!\!\!\!-(\Lie{L_S}{M_S}-\Lie{L_A}{M_A}+\Lie{L_S}{M_A}+\Lie{L_A}{M_S})\Op N_A \\
&&\!\!\!\!+(\Lie{L_S}{N_S}-\Lie{L_A}{N_A}-\Lie{L_S}{N_A}-\Lie{L_A}{N_S})\Op M_S \\
&&\!\!\!\!-(\Lie{L_S}{N_S}-\Lie{L_A}{N_A}+\Lie{L_S}{N_A}+\Lie{L_A}{N_S})\Op M_A\\
&=&\!\!\!\!\LLie{L}{M}\OOp N + M\OOp\LLie{L}{N}\,.
\EAS
\end{Dim}
\end{lemma}

Finally, considering the decomposition of $L \in \Gotico{gl}(n)$ given by $L = L_d + L_p + L_n$
where $L_d$ is diagonal and $L_p$, $L_n$ are respectively the strictly upper and the lower triangular
part of the matrix $M$, it is possible to introduce the map
$$
\begin{array}{ccl}
\rho : \Gotico{gl}(n)&\to&\Gotico{gl}(n)\\
L&\mapsto&L_p - L_n
\end{array}
$$
and the map
$$
\begin{array}{ccl}
\aa : \Gotico{gl}(n)&\to&\Gotico{gl}(n)\\
L&\mapsto&\rho(L_S)
\end{array}
$$
It is well known that the map $\rho$ is a \MR{} on $\Gotico{gl}(n)$ with the standard commutator, moreover it satisfies the following properties:
\begin{lemma}
The map $\rho$ is symmetric (self-adjoint) with respect to the scalar product $\SScal{\cdot}{\cdot}$
and skewsymmetric with respect to the scalar product $\Scal{\cdot}{\cdot}$, i.e.
\BAS
\SScal{L}{\rho(M)} &=& \SScal{\rho(L)}{M}\,, \\
\Scal{L}{\rho(M)} &=& -\Scal{\rho(L)}{M}\,.
\EAS
\begin{Dim}
Using the decomposition $L = L_d + L_p + L_n$, one gets
\begin{eqnarray*}
\SScal{L}{\rho(B)} & = & \Tr(L\rho(M)^t) \\
& = & \Tr[(L_d+L_p+L_n)(M_p-M_n)^t] = \\
& = & \Tr(L_pM_p^t-L_nM_n^t)\,.
\end{eqnarray*}
On the other hand, for the symmetry of the scalar product $((\cdot,\cdot))$, it is
\begin{eqnarray*}
\SScal{\rho(L)}{M} & = & \SScal{M}{\rho(L)} \\
& = & \Tr(M_pL_p^t - M_nL_n^t) = \\
& = & \Tr(L_pM_p^t - L_nM_n^t) \\
&=& \SScal{L}{\rho(M)}\,.
\end{eqnarray*}
In an analogous way, one gets
\begin{eqnarray*}
\Scal{L}{\rho(M)} & = & \Tr(L\rho(M)) \\
& = & \Tr[(L_d+L_p+L_n)(M_p-M_n)] = \\
& = & \Tr(L_nM_p-L_pM_n) \\
&=& -\Scal{\rho(L)}{M}\,.
\end{eqnarray*}
\end{Dim}
\end{lemma}
\begin{lemma}\label{lemmamio2}
The maps $\rho$ and $\aa$ satisfy the relation
$$
\SScal{\rho(L_A)}{M} =\SScal{L}{\aa(M)}\,.
$$
\begin{Dim}
$$
\SScal{\rho(L_A)}{M}= \Scal{\rho(L_A)}{M_S}=-\Scal{L_A}{\rho(M_S)}=\SScal{L}{\aa(M)}\,.
$$
\end{Dim}
\end{lemma}

The vector spaces isomorphism $\Phi$ allows to carry the (restrictions of) Sklyanin and Sem\"enov-Tian-Shansky brackets from $\Gotico{G}^+_{n-1}$ to $\Gotico{gl}(n)$.
\begin{proposition}\label{Skl_bra}
The Sklyanin Poisson bracket carried on $\Gotico{gl}(n)$ through the isomorphism (\ref{isomorfismo}) has the expression:
\BE\label{Pois_Skl}
\begin{array}{rcl}
\ParSkl{f\circ\Phi}{g\circ\Phi}(L_\lambda) &=&
\SScal{L\OOp\rho(L_A)}{\LLie{F}{G}}-\\
&&\SScal{L}{\LLie{F\OOp\rho(L_A)+\aa(L\OOp F)}{G}+\\
&&\LLie{F}{G\OOp\rho(L_A)+\aa(L\OOp G)}}
\end{array}
\EE
where $f$, $g$ are two functions $\Gotico{gl}(n)\to\Reali$  and $F$, $G$ their gradients with respect to $\SScal{\cdot}{\cdot}$.
\end{proposition}
The proof of this statement can be found in the appendix, the corresponding expression for the Sem\"enov-Tian-Shansky bracket on $\Gotico{gl}(n)$ immediately follows from the fact that 
$$
{\Pois{\tilde{f}}{\tilde{g}}}_{Skl}(L_\lambda-\mu\Uno)={\Pois{\tilde{f}}{\tilde{g}}}_{Skl}(L_\lambda)-\mu{\Pois{\tilde{f}}{\tilde{g}}}_{Sem}(L_\lambda)\,.
$$
\begin{proposition}
The Sem\"enov-Tian-Shansky Poisson bracket carried on $\Gotico{gl}(n)$ through the isomorphism (\ref{isomorfismo}) has the expression:
\BE\label{Pois_Sem}
\ParSem{f\circ\Phi}{g\circ\Phi}(L_\lambda)=\SScal{L}{\aa(\LLie{F}{G})-\LLie{\aa(F)}{G}-\LLie{F}{\aa(G)}}\,.
\EE
\begin{Dim}
Substituting $L_\lambda-\mu\Uno$ in (\ref{Pois_Skl}) one obtains
$$
\begin{array}{l}
\SScal{(L\!-\!\mu\Uno)\!\OOp\!\rho(L_A)}{\LLie{F}{G}}\!-\!\SScal{L-\mu\Uno}{\LLie{F\!\OOp\!\rho(L_A)\!+\!\aa[(L\!-\!\mu\Uno)\!\OOp\!F]}{G}\\
+\LLie{F}{G\OOp\rho(L_A)+\aa[(L-\mu\Uno)\OOp G]}} = \\
= \ParSkl{f\!\circ\!\Phi}{g\!\circ\!\Phi}(L_\lambda)\!-\!
\mu\SScal{\rho(L_A)}{\LLie{F}{G}}\!+\!\mu\SScal{L}{\LLie{\aa(F)}{G}\!+\!\LLie{F}{\aa(G)}}
\end{array}
$$
Thus, because
$$
\SScal{\rho(L_A)}{\LLie{F}{G}}=\SScal{L}{\aa(\LLie{F}{G})}
$$
the thesis holds.
\end{Dim}
\end{proposition}

It is now possible to introduce the third bracket of the trihamiltonian structure on $\Gotico{G}^+_{n-1}\sim\Gotico{gl}(n)$: it is given by the formula
\BE\label{Pois_Lin}
\ParLin{f}{g}(L)=\SScal{L}{\LLie{F}{G}}\,.
\EE
\begin{theorem}
The three brackets $\ParSem{\cdot}{\cdot}$, $\ParLin{\cdot}{\cdot}$ and
$\ParSkl{\cdot}{\cdot}$ are mutually compatible Poisson structures, hence they endow
$\Gotico{gl}(n)$ with a trihamiltonian structure.
\begin{Dim}
The two brackets ${\{\cdot,\cdot\}}_{Sem}$ and ${\{\cdot,\cdot\}}_{Skl}$ are compatible Poisson 
brackets on $\Gotico{gl}(n)$ because they are induced by compatible Poisson brackets on
$\Gotico{G}_{n-1}^+\subset\Gotico{G}$. It remains to prove that ${\{\cdot,\cdot\}}_{Lin}$ is a
Poisson bracket and that it is compatible with both the others.

The fact that it is a Poisson bracket follows immediately from Lemma \ref{Lie doppia} and the fact that it is of Lie--Poisson type. On the other hand considering the two vector field
\BAS
X(L) &=& \rho(L_A) \\
Y(L) &=& L\OOp\rho(L_A)
\EAS
one observes that
\BAS
\diff{X}\cdot M &=& \left.\frac{\D}{\D \tau}\right|_{\tau=0}\!\!\!\rho(L_A+\tau M_A) \\
&=& \rho(M_A)
\EAS
\BAS
\SScal{\diff{X}\cdot M}{N} &=& \SScal{\rho(M_A)}{N}\\
&=& \SScal{M}{\aa(N)}
\EAS
and
\BAS
\diff{Y}\cdot M &=& \left.\frac{\D}{\D \tau}\right|_{\tau=0}\!\!\!(L+\tau M)\OOp\rho(L_A+\tau M_A) \\
&=& L\OOp\rho(M_A)+M\OOp\rho(L_A)
\EAS
\BAS
\SScal{\diff{Y}\cdot M}{N} &=& \SScal{L\OOp\rho(M_A)+M\OOp\rho(L_A)}{N}\\
&=&\SScal{M}{N\OOp\rho(L_A)}+\SScal{\rho(M_A)}{L\OOp N}\\
&=& \SScal{M}{N\OOp\rho(L_A)+\aa(L\OOp N)}\,.
\EAS
Hence $\Tras{\diff{X}}\cdot N= \aa(N)$ and $\Tras{\diff{Y}}\cdot N= N\OOp\rho(L_A)+\aa(L\OOp N)$. Lemma~\ref{derivo} and Lemma~\ref{lemmamio2} imply that both the Sem\"enov-Tian-Shansky and the Sklyanin tensors are the Lie derivative of the Poisson tensor
associated to ${\{\cdot,\cdot\}}_{Lin}$, respectively trough the vector fields $X$ and $Y$. The two structures $\ParSem{\cdot}{\cdot}$ and $\ParSkl{\cdot}{\cdot}$ are thus compatible with the new structure $\ParLin{\cdot}{\cdot}$.
\end{Dim}
\end{theorem}

\section{Application to periodic Toda lattice}
The aim of this section is to apply the trihamiltonian structure given in the previous section to the construction of a trihamiltonian extension of the periodic Toda lattice. The extended system has a richer geometrical structure than the ordinary one: indeed, only two of the three Poisson structures can be reduced on the subspace on which the ordinary periodic Toda lattice is defined, producing its well known bihamiltonian formulation, the new linear structure (\ref{Pois_Lin}) is not reducible.

The periodic $n$-particles Toda lattice has been extensively investigated and a wide literature has been produced on this subject. Here just few basic facts are recalled in order to fix the notation and to allow the application of the general trihamiltonian construction. The exposition is largely based on the paper~\cite{MorPiz1996a} to which the reader is referred for details and proofs; see also~\cite{ORagn1989} and~\cite{tesiChiara}.

\subsection{Trihamiltonian extension}
The first step in the study of the Toda lattice is usually the introduction of the so-called Flaschka coordinates
\BAS
a_i &=& \frac{1}{2} e^{\frac{1}{2}(x_i-x_{i+1})} \,, \\
b_i &=& - \frac{1}{2} p_i \,.
\EAS
In these coordinates the evolution equations for the periodic Toda lattice become
\BE\label{eqsFlaschkaper}
\dot{a}_i = a_i(b_{i+1}-b_i) \quad \dot{b}_i = 2(a_i^2-a_{i-1}^2) \quad i=1 \ldots n
\EE
where, as usual, one has the periodicity condition $a_{i+n}=a_i$, $b_{i+n}=b_i$.

Choosing the Lax pair
\BE\label{LaxMat}
\begin{array}{l}
L_\lambda=\left(
\begin{array}{cccccc}
b_1 & a_1 & 0 & \cdots & a_n/\lambda \\
a_1 & b_2 & \ddots & \ddots & \vdots \\
\vdots & \ddots & \ddots & \ddots& 0 \\
0 && \ddots & \ddots & a_{n-1} \\
a_n\lambda & 0 & \cdots &a_{n-1} & b_n
\end{array}
\right)\\[1.5cm]
B_\lambda=\left(
\begin{array}{cccccc}
0 & a_1 & 0 & \cdots & -a_n/\lambda \\
-a_1 & 0 & \ddots & \ddots & \vdots \\
\vdots & \ddots & \ddots & \ddots& 0 \\
0 && \ddots & \ddots & a_{n-1} \\
a_n\lambda & 0 & \cdots &-a_{n-1} & 0
\end{array}
\right)
\end{array}
\EE
the equations (\ref{eqsFlaschkaper}) can be put in Lax form (with a spectral parameter):
\BE\label{Lax}
\frac{\D}{\D t}L_\lambda=\Lie{L_\lambda}{B_\lambda}\,.
\EE
The matrix $L_\lambda$ in equation (\ref{Lax}) is easily recognized as an element of the subspace $\Gotico{G}_1^+$, with $\sigma_0=(b_1,\ldots,b_n)$ and $\sigma_1=(a_1,\ldots,a_n)$. The Poisson tensors $\widehat{P}_{Sem}$ and  $\widehat{P}_{Skl}$, obtained from the reduction of the two structure (\ref{Sem1}) and (\ref{Skl1}) on $\Gotico{G}^+_1$, allow to write the vector field corresponding to Lax equations (\ref{Lax}) in a bihamiltonian way:
\BE\label{Todavet}
X_{Toda}=\widehat{P}_{Sem}\D \widehat{H}_{Sem}=\widehat{P}_{Skl}\D \widehat{H}_{Skl}
\EE
where the Hamiltonians are given by
$$
\widehat{H}_{Skl}=\TrKup(L_\lambda),\quad \widehat{H}_{Sem}=\frac{\TrKup(L_\lambda^2)-(\TrKup(L_\lambda))^2}{2}\,.
$$
Taken two Hamiltonians $H_{Sem}$ and $H_{Skl}$, on $\Gotico{G}_k^+\supset \Gotico{G}_1^+$, such that
\BE\label{eqrid}
H_{Sem}|_{\Gotico{G}_1}=\widehat{H}_{Sem}, \quad H_{Skl}|_{\Gotico{G}_1}=\widehat{H}_{Skl}
\EE
the bihamiltonian vector field associated to them by the (suitable reduced) Poisson structures (\ref{Sem1}) and (\ref{Skl1}) can be thought as an extension of the periodic Toda lattice.

The following theorem states that there exist an extension $X_{Ext}$ of Toda vector field $X_{Toda}$ that is trihamiltonian with respect to the three structure $\ParSem{\cdot}{\cdot}$, $\ParLin{\cdot}{\cdot}$ and $\ParSkl{\cdot}{\cdot}$.
\begin{theorem}\label{hamiltoniane3}
Indicated with $\PSem{}$, $\PLin{}$ and $\PSkl{}$ the Poisson tensors corresponding respectively to the brackets $\ParSem{\cdot}{\cdot}$, $\ParLin{\cdot}{\cdot}$ and $\ParSkl{\cdot}{\cdot}$, the recurrence relations:
$$
\PSem{} \D H_{Sem} =\PLin{} \D H_{Lin} =\PSkl{} \D H_{Skl}
$$
hold, where
\BAS
H_{Skl} &=& \Tr(L)\,,\\
H_{Lin} &=&\Tr(L_A\Op\aa(L))\,,\\
H_{Sem} &=& \frac{\Tr(L_S^2)-\Tr(L_A^2)-(\Tr(L_S))^2}{2}\,.
\EAS
The vector field $X_{Ext}=\PSem{} \D H_{Sem}$ is an extension of the Toda vector field (\ref{Todavet}) in the sense of equations (\ref{eqrid}).
\begin{Dim}
Firstly the gradient of $H_{Skl}$, $H_{Lin}$ and $H_{Sem}$ with respect to the scalar product
$\SScal{\cdot}{\cdot}$ are needed. The simple calculations
\BAS
\SScal{\nabla \Tr(L)}{M} &=& \left.\frac{\D}{\D \tau}\right|_{\tau=0}\hspace{-7pt}\Tr(L+\tau M) \\
&=&\Scal{M}{\Uno^t} \\
&=& \SScal{M}{\Uno}\\
\SScal{\Tr(L_A\Op\aa(L)))}{M} &=& \left.\frac{\D}{\D \tau}\right|_{\tau=0}\hspace{-7pt}\Tr\big[(L_A+\tau M_A)\Op\aa(L+\tau
M)\big]\\ 
&=& \Scal{L_A}{\aa(M)}+\Scal{M_A}{\aa(L)} \\
&=&  \Scal{L_A}{\rho(M_S)}+\Scal{M_A}{\rho(L_S)} \\
&=& \Scal{\rho(L_S)}{M_A} - \Scal{M_S}{\rho(L_A)} \\
&=& -\SScal{\rho(L)}{M} \\
\SScal{\nabla H_{Sem}}{M} &=& \left.\frac{\D}{\D \tau}\right|_{\tau=0}\hspace{-7pt}\frac{\Tr\left[{(L_S\!+\!\tau
M_s)}^2\right]\!-\!\Tr\left[{(L_A\!+\!\tau M_A)}^2\right]}{2} \\
&&-\left.\frac{\D}{\D \tau}\right|_{\tau=0}\hspace{-7pt}\frac{{\left[\Tr(L_S+\tau M_S)\right]}^2}{2}\\
&=& \Scal{L_S}{M_S}-\Scal{L_A}{M_A}-\Scal{\Tr(L_S)\Uno}{M_S} \\
&=& \SScal{L-\Tr(L)\Uno}{M}
\EAS
give
\BAS
\nabla H_{Skl} &=& \Uno\,, \\
\nabla H_{Lin} &=& -\rho(L)\,, \\
\nabla H_{Sem} &=& L-\Tr(L)\Uno\,.
\EAS
Then substituting the previous expressions for the gradient respectively in (\ref{Pois_Lin}), (\ref{Pois_Skl}) and (\ref{Pois_Sem}) one obtains, for any function $g$:
\BAS
\ParLin{H_{_{Lin}}}{g} &=& -\SScal{L}{\LLie{\rho(L)}{\nabla g}} \,,\\
\ParSkl{H_{_{Skl}}}{g} &=& -\SScal{L}{\LLie{\Uno\OOp\rho(L_A)+\aa(L\OOp\Uno)}{\nabla g}}\\
&=& -\SScal{L}{\LLie{\rho(L_A)+\rho(L_S)}{\nabla g}} \\
&=& -\SScal{L}{\LLie{\rho(L)}{\nabla g}} \,,\\
\ParSem{H_{_{Sem}}}{g} &=& \SScal{L}{\aa(\LLie{L}{\nabla g})-\LLie{\aa(L)}{\nabla
g}-\LLie{L}{\aa(\nabla g)}} \\
&=& \SScal{\rho(L_A)}{\LLie{L}{G}}-\SScal{\rho(L_S)}{\LLie{G}{L}} \\
&=& -\SScal{L}{\LLie{\rho(L)}{G}}\,.
\EAS
Thus the recursion relation holds. The identities (\ref{eqrid}) are easily proved: because $L_\lambda=L_0+L_1\lambda+L_1^t\lambda^{-1}$ both on $\Gotico{G}_{n-1}$ and $\Gotico{G}_1$ one has immediately $\TrKup(L_\lambda)=\Tr(L_0)=\Tr(L)$; moreover
\BAS
\TrKup(L_\lambda^2)-(\TrKup(L_\lambda))^2 &=&\Tr(L_0^2)-(\Tr(L_0))^2+2\Tr(L_1L_1^t) \\
&=& \Tr(L_S^2)-(\Tr(L_S))^2-\Tr(L_A^2)
\EAS
because $\Tr(L_A^2)=\Tr((L_1-L_1^t)^2)=-2\Tr(L_1L_1^t)$, being $L_1$ a strictly triangular matrix.
\end{Dim}
\end{theorem}
A final observation on the choice of the Hamiltonians is necessary: it is also possible to obtain the vector fields $X_{Ext}$ and $X_{Toda}$ respectively with the two alternative Hamiltonians
$$
H_{Sem} = \frac{\Tr(L_S^2)-\Tr(L_A^2)}{2}, \quad \widehat{H}_{Sem}=\frac{1}{2}\TrKup(L_\lambda^2)
$$
but the choice of Theorem~\ref{hamiltoniane3}, as will be shown in the examples, allows to find a recurrence scheme with a finite number of Hamiltonians.
\subsection{Some examples}
The ordinary periodic Toda lattice with $3$ particles is set in the subspace $\Gotico{G}_1^+$ of $\Gotico{gl}
(3)(\lambda, \lambda^{-1}]$. In Flaschka coordinates, $(b_i,a_i)$ the formula for the Lax 
matrix (\ref{LaxMat}), becomes 
$$
L=\Delta^{-1}a+b+a\Delta \equiv \left(
\begin{array}{ccc}
b_1 & \;\;a_1 & \;a_3/\lambda \\
a_1 & \;\;b_2 & a_2 \\
a_3\lambda & \;\;a_2 & b_3
\end{array}
\right)\,.
$$
The extension is obtained considering the subspace $\Gotico{G}_2^+$ with coordinates $(b_i, a_i, c_i)$, corresponding to a Lax matrix
$$
L=\Delta^{-2}c+\Delta^{-1}a+b+a\Delta+c\Delta^2 \equiv \left(
\begin{array}{ccc}
b_1 & a_1+\displaystyle{\frac{c_1}{\lambda}}& \;\;c_3+\displaystyle{\frac{a_3}{\lambda}} \\
a_1+c_1\lambda & b_2 & a_2+\displaystyle{\frac{c_2}{\lambda}} \\
c_3+a_3\lambda & \;\;a_2+c_2\lambda & b_3
\end{array}
\right)\,.
$$

The three Poisson structures $\PSem$, $\PSkl$ and $\PLin$, given respectively by (\ref{Sem1}), (\ref{Skl1}) and (\ref{Pois_Lin}), are
$$
\PSem = \\
 \left(
\begin {array}{ccccccccc}
0  & 0 & 0 & a_1 & 0 & -a_3 & c_1 & c_2 & 0 \\
\noalign{\medskip}
& 0 & 0 & -a_1 & a_2 & 0 & 0 & c_2 & -c_3 \\
\noalign{\medskip}
&& 0 & 0 & -a_2 & a_3 & -c_1 & 0 & c_3 \\
\noalign{\medskip}
&&& 0 & c_1 & -c_3 & 0 & 0 & 0 \\
\noalign{\medskip}
&&&& 0 & c_2 & 0 & 0 & 0 \\
\noalign{\medskip}
&&&&& 0 & 0 & 0 & 0 \\
\noalign{\medskip}
&&&\ast&&& 0 & 0 & 0 \\
\noalign{\medskip}
&&&&&&& 0 & 0 \\
\noalign{\medskip}
&&&&&&&& 0
\end {array} \right)
$$
$$
\PSkl =\left(\tiny{
\begin {array}{ccccccccc}
\hspace{-7pt}0 &\hspace{-7pt} 2(a_1^2\!-\!c_2^2) &\hspace{-7pt} 2(c_1^2\!-\!a_3^2) & \hspace{-7pt}b_1a_1 &\hspace{-12pt} 2(c_1a_1\!-\!a_3c_2) & \hspace{-7pt}-b_1a_3
&\hspace{-10pt} b_1c_1 &\hspace{-10pt} -b_1c_2 & 0 \\
\noalign{\medskip}
& 0 &\hspace{-7pt} 2(a_2^2\!-\!c_3^2) & \hspace{-7pt}-b_2a_1 & b_2a_2 &\hspace{-10pt} 2(c_2a_2\!-\!c_3a_1)
& 0 &\hspace{-10pt} b_2c_2 &\hspace{-10pt} -b_2c_3 \\
\noalign{\medskip}
& & 0&\hspace{-10pt} 2(a_3c_3\!-\!c_1a_2) & -b_3a_2 & b_3a_3
&\hspace{-10pt} -b_3c_1 & 0 &\hspace{-10pt} b_3c_3 \\
\noalign{\medskip}
& & & 0 &\hspace{-12pt}  \begin{array}{c}\frac{a_2a_1}{2}+\\ c_1b_2\!-\!c_3c_2\end{array} &\hspace{-10pt} \begin{array}{c}-\frac{a_2a_3}{2}+\\ c_1c_2\!-\!b_1c_3 \end{array}
&\hspace{-10pt} \frac{a_1c_1}{2} & 0 &\hspace{-10pt} -\frac{a_1c_3}{2} \\
\noalign{\medskip}
& & & & 0 &\hspace{-10pt} \begin{array}{c}\frac{a_2a_3}{2}+\\ b_3c_2\!-\!c_1c_3 \end{array}
&\hspace{-10pt} -\frac{a_2c_1}{2} &\hspace{-10pt} \frac{a_2c_2}{2} & 0 \\
\noalign{\medskip}
& & \ast & & & 0 & 0 &\hspace{-10pt} -\frac{a_3c_2}{2} &\hspace{-10pt} \frac{a_3c_3}{2} \\
\noalign{\medskip}
& & & & &  & 0 &\hspace{-10pt} -\frac{c_1c_2}{2} &\hspace{-10pt} \frac{c_1c_3}{2} \\
\noalign{\medskip}
& & & & & & & 0 &\hspace{-10pt} -\frac{c_2c_3}{2} \\
\noalign{\medskip}
& & & & & & & & 0
\end{array}
}\right)
$$
$$
\PLin = 
\left(
\begin{array}{ccccccccc}
0 & 0 & 0 & -2c_2 & 0 & 2c_1& -2a_3 & 2a_1 & 0 \\
\noalign{\medskip}
& 0 & 0 & c_2 & -2c_3 & 0 & 0 & -2a_1 & 2a_2 \\
\noalign{\medskip}
& & 0 & 0 & 2c_3 & -2c_1 & 2a_3 & 0 & -2a_2\\
\noalign{\medskip}
& & & 0& -a_3 & a_2 & -c_3 & b_2-b_1 & c_1 \\
\noalign{\medskip}
& & & & 0 & -a_1 & c_2 & -c_1 & b_3-b_2 \\
\noalign{\medskip}
& & & & & 0 & b_1-b_3 & c_3 & -c_2 \\
\noalign{\medskip}
& & & \ast & & & 0 & a_2 & -a_1 \\
\noalign{\medskip}
& & & & & & & 0 & a_3 \\
\noalign{\medskip}
& & & & & & & & 0
\end {array} \right)
$$

The Hamiltonians satisfying Theorem~\ref{hamiltoniane3} are in this case
\begin{eqnarray*}
H_{Skl} = H_{13} & = & b_1 + b_2 + b_3\,, \\
H_{Lin} = H_{22} & = & a_1 c_2 + a_2 c_3 + a_3 c_1 \,,\\
H_{Sem} = H_{12} & = & a_1^2 + a_2^2 + a_3^2 - b_1 b_2 - b_1 b_3 - b_2 b_3\,.\\
&&+ c_1^2 + c_2^2 + c_3^2
\end{eqnarray*}

Moreover, using all the three Poisson structure, it is possible to construct the following recursion scheme, that prove the complete in\-te\-gra\-bi\-li\-ty of the system:
\begin{center}
\begin{picture}(150,165)
\put(30,0){\BloccoS{}{0}}
\put(75,0){\BloccoS{$H_{12}$}{$0$}} \put(120,0){\BloccoS{}{$0$}}
\put(0,35){\BloccoQ{$H_{11}-2H_{31}$}{$0$}} \put(45,50){\MiniNucleoA}
\put(90,50){\MiniNucleoA} \put(120,35){\BloccoP{$H_{13}$}{$0$}}
\put(22.5,80){\BloccoQ{$H_{21}$}{$0$}} \put(67.5,95){\MiniNucleoA}
\put(97.5,80){\BloccoP{$H_{22}$}{$0$}}
\put(45,125){\BloccoQ{$H_{31}$}{$0$}} \put(75,125){\BloccoP{}{$0$}}
\end{picture}
\end{center}
where $P=\PSem$, $Q=\PSkl$, $S=\PLin$ and
\begin{eqnarray*}
H_{11} & = & b_1b_2b_3 -b_1(a_2^2+c_3^2) -b_2(a_3^2+c_1^2) -b_3(a_1^2+c_2^2) \\
&& -2a_1a_2c_1 -2a_2a_3c_2 -2a_3a_1c_3 - 2c_1c_2c_3 \,,\\
H_{21} & = & a_1a_2a_3 +a_1(c_1c_3-c_2b_3) +a_2(c_2c_1-c_3b_1) +a_3(c_3c_2-c_1b_2) \,,\\
H_{31} & = & c_1c_2c_3\,.
\end{eqnarray*}

Analogous results can be obtained also for higher dimensions: in the cases $n=4$ and $n=5$, by direct computation, one can respectively construct the recursion schemes 
$$
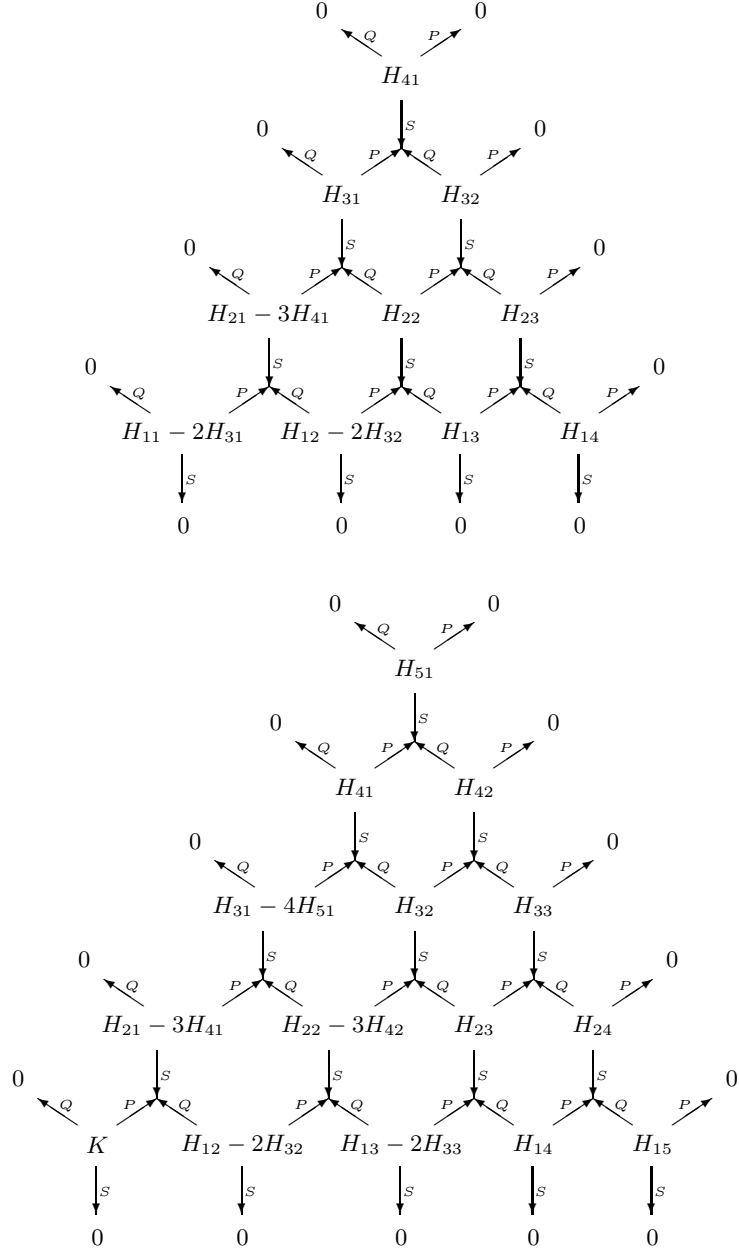
\begin{picture}(220,220)
\put(25,0){\BloccoS{$H_{11}-2H_{31}$}{$0$}}
\put(85,0){\BloccoS{$H_{12}-2H_{32}$}{$0$}}
\put(130,0){\BloccoS{$H_{13}$}{$0$}}
\put(175,0){\BloccoS{$H_{14}$}{$0$}}
\put(-10,35){\BloccoQ{}{$0$}}
\put(50,50){\MiniNucleoA}
\put(100,50){\MiniNucleoA}
\put(145,50){\MiniNucleoA}
\put(175,35){\BloccoP{}{$0$}}
\put(27.5,80){\BloccoQ{$H_{21}-3H_{41}$}{$0$}}
\put(77.5,95){\MiniNucleoA}
\put(107.5,80){\Scatola{$H_{22}$}}
\put(122.5,95){\MiniNucleoA}
\put(152.5,80){\BloccoP{$H_{23}$}{$0$}}
\put(55,125){\BloccoQ{$H_{31}$}{$0$}}
\put(100,140){\MiniNucleoA}
\put(130,125){\BloccoP{$H_{32}$}{$0$}}
\put(77.5,170){\BloccoQ{$H_{41}$}{$0$}}
\put(107.5,170){\BloccoP{}{$0$}}
\end{picture}
$$
$$
\begin{picture}(350,260)
\put(30,0){\BloccoS{$K$}{$0$}}
\put(85,0){\BloccoS{$H_{12}-2H_{32}$}{$0$}}
\put(145,0){\BloccoS{$H_{13}-2H_{33}$}{$0$}}
\put(195,0){\BloccoS{$H_{14}$}{$0$}}
\put(240,0){\BloccoS{$H_{15}$}{$0$}}
\put(0,35){\BloccoQ{}{$0$}}
\put(45,50){\MiniNucleoA}
\put(110,50){\MiniNucleoA}
\put(165,50){\MiniNucleoA}
\put(210,50){\MiniNucleoA}
\put(240,35){\BloccoP{}{$0$}}
\put(25,80){\BloccoQ{$H_{21}-3H_{41}$}{$0$}}
\put(85,95){\MiniNucleoA}
\put(123,80){\Scatola{$H_{22}-3H_{42}$}}
\put(142.5,95){\MiniNucleoA}
\put(172.5,80){\Scatola{$H_{23}$}}
\put(187.5,95){\MiniNucleoA}
\put(217.5,80){\BloccoP{$H_{24}$}{$0$}}
\put(67,125){\BloccoQ{$H_{31}-4H_{51}$}{$0$}}
\put(120,140){\MiniNucleoA}
\put(150,125){\Scatola{$H_{32}$}}
\put(165,140){\MiniNucleoA}
\put(195,125){\BloccoP{$H_{33}$}{$0$}}
\put(97.5,170){\BloccoQ{$H_{41}$}{$0$}}
\put(142.5,185){\MiniNucleoA}
\put(172.5,170){\BloccoP{$H_{42}$}{$0$}}
\put(120,215){\BloccoQ{$H_{51}$}{$0$}}
\put(150,215){\BloccoP{}{$0$}}
\end{picture}
$$
Where
\BAS
&&\det(L_\lambda-\mu)=(-\mu)^n+H_1(\lambda+\lambda^{-1})+\cdots+H_n(\lambda^n+\lambda{-n})\\
&&H_i=\sum_{j=1}^{n-i+1}H_{ij}\mu^j\,,\qquad K=H_{11}-2H_{31}+2H_{51}
\EAS
for $L_\lambda\in\Gotico{G}_{n-1}^+\subset\Gotico{gl}(n)(\lambda, \lambda^{-1}]$.

The regularity in the schemes leads naturally to believe that the construction is possible for all $n$, i.e. to conjecture that the extension of the periodic Toda lattice with $n$ particles introduced in this paper is not only trihamiltonian, but also completely integrable.

\section{Conclusions}
In this paper a trihamiltonian structure is constructed on a subspace of the Kupershmidt algebra of shift operators on $n$-periodic sequences. This subspace is formed by symmetric operators of maximum degree $n-1$ and it is isomorphic to $\Gotico{gl}(n)$. The first two compatible Poisson bracket that define the trihamiltonian structure are the well known Sem\"enov-Tian-Shansky and Sklyanin brackets obtained from \MR techniques. The third Poisson bracket is new: it is the Lie--Poisson structure associated to a non-standard Lie bracket on $\Gotico{gl}(n)$. As an application an trihamiltonian extension of the periodic Toda lattice with $n$ particles is constructed, and a complete set of first integrals mutually in involution for the extended system up to five particles is found using a trihamiltonian recursion scheme.

\section*{Acknowledgements}
This work has been partially supported by the PRIN research project ``\textit{Geometric methods in the theory of nonlinear waves and their applications}'' of the Italian MIUR.

\newpage
\section*{Appendix}
In this appendix the expression (\ref{Pois_Skl}) appearing in Proposition~\ref{Skl_bra} is proved. Set $\tilde{f}=f \circ \Phi$ and $M=\Phi(M_\lambda)$, one can check the identity
$$
\Scal{\nabla\tilde{f}}{M_\lambda} = \SScal{\nabla f}{M} \quad \forall M_\lambda\in\Gotico{G}_{n-1}^+\,.
$$
This first observation implies that if $F=\nabla f=F_S+F_A$ then
$$
\nabla\tilde{f}=\Phi^{-1}(F)=F_0+F_1\lambda+F_1^t\lambda^{-1}=F_\lambda
$$
with $F_S=F_0$ and $F_A=F_1-F_1^t$. The second observation is that if $M\in\Gotico{G}_+$ (i.e. $\Adg M=-M$ holds), then
$$
M^+=\rho(M_0)+\sum_{k>0}M_k\lambda^k + \sum_{k>0}M_k^t\lambda^{-k}
$$
while if $M\in\Gotico{G}^+$ (i.e. $\Adg M=M$ holds) then
$$
M_+=\rho(M_0)+\sum_{k>0}M_k\lambda^k - \sum_{k>0}M_k^t\lambda^{-k}\,.
$$
These two facts follows immediately from the definitions of the two projections on $\Gotico{G}_+$ and $\Gotico{G}^+$. Finally it is worth noting that the property (\ref{der_prop}) implies that
\BE\label{cons_der1}
\begin{array}{rcl}
\Lie{A}{B\Op C}-A\Op\Lie{B}{C} &=& -\Lie{B}{A}\Op C + B\Op\Lie{A}{C}-A\Op\Lie{B}{C} \\
&=& -\Lie{B}{A\Op C}+B\Op\Lie{A}{C}
\end{array}
\EE
and moreover
\BE\label{cons_der2}
\begin{array}{rcl}
\Lie{A}{B\Op C}-A\Op\Lie{B}{C} &=& \Lie{A}{B\Op C}+\Lie{A}{A\Op C}\\
&&-A\Op\Lie{A}{C}-A\Op\Lie{B}{C} \\
&=&\Lie{A}{(A+B)\Op C}-A\Op\Lie{A+B}{C}\,.
\end{array}
\EE

Using the representation (\ref{CampiSkl}) for an Hamiltonian vector field for the Sklyanin bracket one has
\BAS
\ParSkl{\tilde{f}}{\tilde{g}} = \Scal{G_\lambda}{\XSkl{\tilde{f}}} &=& \SScal{G}{\Phi\left(\Lie{L_\lambda}{(L_\lambda\Op F_\lambda)_+}-L_\lambda\Op\Lie{L_\lambda}{F_\lambda}^+\right)}
\EAS
and thus it is essential to calculate $\Phi\left(\Lie{L_\lambda}{(L_\lambda\Op F_\lambda)_+}-L_\lambda\Op\Lie{L_\lambda}{F_\lambda}^+\right)$. A tedious computation gives
\BAS
L_\lambda\Op\Lie{L_\lambda}{F_\lambda}^+ \!\!\!\!&=&\!\!\!\!
L_0\Op\rho(\Lie{L_0}{F_0}+\Lie{L_1}{F_1^t}+\Lie{L_1^t}{F_1}) +\\
&&\!\!\!\! L_1^t\Op(\Lie{L_0}{F_1}\!+\!\Lie{L_1}{F_0})-L_1\Op(\Lie{L_0}{F_1^t}\!+\!\Lie{L_1^t}{F_0})+ \\
&&\!\!\!\! \lambda\left[
L_0\Op(\Lie{L_0}{F_1}+\Lie{L_1}{F_0})+
L_1^t\Op\Lie{L_1}{F_1}+\right.\\
&&\!\!\!\! \left.L_1\Op\rho(\Lie{L_0}{F_0}+\Lie{L_1}{F_1^t}+\Lie{L_1^t}{F_1})
\right]+\\
&&\!\!\!\! \lambda^2\left[
L_0\Op\Lie{L_1}{F_1}+L_1\Op(\Lie{L_0}{F_1}+\Lie{L_1}{F_0})
\right]+\\
&&\!\!\!\! \lambda^3L_1\Op\Lie{L_1}{F_1}\\
&&\!\!\!\! +\lambda^{-1}{(\cdots)}^t+\lambda^{-2}{(\cdots)}^t+\lambda^{-3}{(\cdots)}^t\\[10pt]
\Lie{L_\lambda}{(L_\lambda\Op F_\lambda)_+} \!\!\!\!&=&\!\!\!\!
\Lie{L_0}{\rho(L_0\Op F_0+L_1\Op F_1^t+L_1^t\Op F_1)} +\\
&&\!\!\!\! \Lie{L_1^t}{L_0\Op F_1+L_1\Op F_0}-
\Lie{L_1}{L_0\Op F_1^t+L_1^t\Op F_0}+ \\
&&\!\!\!\! \lambda\left(
\Lie{L_0}{L_0\Op F_1+L_1\Op F_0}+
\Lie{L_1^t}{L_1\Op F_1}+\right.\\
&&\!\!\!\! \left.\Lie{L_1}{\rho(L_0\Op F_0+\L_1\Op F_1^t+L_1^t\Op F_1)}
\right)+\\
&&\!\!\!\! \lambda^2\left(
\Lie{L_0}{L_1\Op F_1}+\Lie{L_1}{L_0\Op F_1+L_1\Op F_0}
\right)+\\
&&\!\!\!\! \lambda^3\Lie{L_1}{L_1\Op F_1}\\
&&\!\!\!\! +\lambda^{-1}{(\cdots)}^t+\lambda^{-2}{(\cdots)}^t+\lambda^{-3}{(\cdots)}^t
\EAS
where the coefficients of the negative powers of $\lambda$ are the transposed of the coefficient of the positive powers. As expected from the reduction theorem, in the two previous terms the coefficients of the powers of $\lambda$ grater than one are equal. Thus using the identities
\BAS
&&L_1\Op F_1^t+L_1^t\Op F_1 = L_1\Op F_1 + L_1^t\Op F_1^t-(L_1-L_1^t)\Op(F_1-F_1^t)\\
&&\Lie{L_1}{F_1^t}+\Lie{L_1^t}{F_1} = \Lie{L_1}{F_1} + \Lie{L_1^t}{F_1^t}-\Lie{L_1-L_1^t}{F_1-F_1^t}\\
&&\rho(L_1\Op F_1)= -L_1\Op F_1\,,\quad \rho(\Lie{L_1}{F_1})= -\Lie{L_1}{F_1}\\
&&\rho(L_1^t\Op F_1^t)= -L_1^t\Op F_1^t\,,\quad \rho(\Lie{L_1^t}{F_1^t})= -\Lie{L_1^t}{F_1^t}
\EAS
one obtains the following expression:
\BAS
&&\Lie{L_\lambda}{(L_\lambda\Op F_\lambda)_+}-L_\lambda\Op\Lie{L_\lambda}{F_\lambda}^+=\\
&=& \Lie{L_0}{\rho(L_S\Op F_S-L_A\Op F_A)}-\Lie{L_0}{L_1\Op F_1-L_1^t\Op F_1^t}-\\
&&L_0\Op\rho(\Lie{L_S}{F_S}-\Lie{L_A}{F_A})+L_0\Op(\Lie{L_1}{F_1}-\Lie{L_1^t}{F_1^t})- \\
&&\Lie{L_0}{F_1\Op L_1^t}+\Lie{L_0}{F_1^t\Op L_1}-
L_0\Op\Lie{F_1}{L_1^t}+L_0\Op\Lie{F_1^t}{L_1}+\\
&&\Lie{L_1^t}{L_1\Op F_0}-L_1^t\Op\Lie{L_1}{F_0}-
\Lie{L_1}{L_1^t\Op F_0}+L_1\Op\Lie{L_1^t}{F_0}+\\
&&\lambda\Big(
\Lie{L_1}{\rho(L_S\Op F_S-L_A\Op F_A)}-L_1\Op\rho(\Lie{L_S}{F_S}-\Lie{L_A}{F_A})+\\
&&\Lie{L_1^t}{L_1\Op F_1}-L_1^t\Op\Lie{L_1}{F_1}+
\Lie{L_1}{L_1^t\Op F_1^t}-L_1\Op\Lie{L_1^t}{F_1^t}+\\
&&\Lie{L_0}{L_1\Op F_0}-L_0\Op\Lie{L_1}{F_0}
\Big)+\lambda^{-1}{(\cdots)}^t=
\EAS
using the properties (\ref{cons_der1}), (\ref{cons_der2}) and the fact that $L_1+L_1^t=-\rho(L_A)$
\BAS
&\!\!\!\! =&\!\!\!\! \Lie{L_S}{\rho(L_S\Op F_S-L_A\Op F_A)}-L_S\Op\rho(\Lie{L_S}{F_S}-\Lie{L_A}{F_A})+\\
&&\!\!\!\! \Lie{L_S}{\rho(L_A)\!\Op\!F_A}\!-\!L_S\!\Op\!\Lie{\rho(L_A)}{F_A}\!+\!\Lie{L_A}{\rho(L_A)\!\Op\!F_S}\!-\!L_A\!\Op\!\Lie{\rho(L_A)}{ F_S}\!+\\
&&\!\!\!\! \lambda\Big(
\Lie{L_1}{\rho(L_S\Op F_S-L_A\Op F_A)}-
L_1\Op\rho(\Lie{L_S}{F_S}-\Lie{L_A}{F_A})+\\
&&\!\!\!\! \Lie{L_1^t}{L_1\Op F_1}-L_1^t\Op\Lie{L_1}{F_1}+
\Lie{L_1}{L_1^t\Op F_1^t}-L_1\Op\Lie{L_1^t}{F_1^t}+\\
&&\!\!\!\! \Lie{L_0}{L_1\Op F_0}-L_0\Op\Lie{L_1}{F_0}
\Big)-\lambda^{-1}\Big(
\Lie{L_0}{L_1^t\Op F_0}-L_0\Op\Lie{L_1^t}{F_0}+\\
&&\!\!\!\! L_1^t\Op\rho(\Lie{L_S}{F_S}-\Lie{L_A}{F_A})-\Lie{L_1^t}{\rho(L_S\Op F_S-L_A\Op F_A)}+\\
&&\!\!\!\! L_1\Op\Lie{L_1^t}{F_1}-\Lie{L_1}{L_1^t\Op F_1}+
L_1^t\Op\Lie{L_1}{F_1^t}-\Lie{L_1^t}{L_1\Op F_1^t}
\Big)\,.
\EAS
Finally, applying $\Phi$, the quantity
$$
\begin{array}{ll}
\Scal{G_\lambda}{\Lie{L_\lambda}{(L_\lambda\Op F_\lambda)_+}\;-&\!\!\!\!\!L_\lambda\Op\Lie{L_\lambda}{F_\lambda}^+}=\\
&=\SScal{G}{\Phi\left(\Lie{L_\lambda}{(L_\lambda\Op F_\lambda)_+}-L_\lambda\Op\Lie{L_\lambda}{F_\lambda}^+\right)}
\end{array}
$$
becomes
\BAS
&&\!\!\!\! \bigg(G_S,
\Lie{L_S}{\aa(L\OOp F)}-L_S\Op\rho({\LLie{L}{F}}_A)+\Lie{L_S}{\rho(L_A)\Op F_A}\\
&&\!\!\!\! +\Lie{L_A}{\rho(L_A)\Op F_S}+L_S\Op{\LLie{\rho(L_A)}{F}}_S-L_A\Op{\LLie{\rho(L_A)}{F}}_A
\bigg)-\\
&&\!\!\!\! \bigg(G_A,
L_S\Op{\LLie{\rho(L_A)}{F}}_A+L_A\Op{\LLie{\rho(L_A)}{F_A}}_S+\Lie{L_A}{\aa(L\OOp F)}\\
&&\!\!\!\! -L_A\Op\rho({\LLie{L}{F}}_A)+\Lie{L_A}{\rho(L_A)\Op F_A}-\Lie{L_S}{\rho(L_A)\Op F_S}
\bigg)\\
&\!\!\!\!=&\!\!\!\!
\SScal{G}{L\!\OOp\!\LLie{\rho(L_A)}{F}-\LLie{L}{\rho(L_A)\!\OOp\!F}-\LLie{L}{\aa(L\!\OOp\!F}-L\!\OOp\!\rho({\LLie{L}{F}}_A)}\\
&\!\!\!\!=&\!\!\!\! \SScal{G}{\LLie{L\OOp\rho(L_A)}{F}-\rho(L_A)\OOp\LLie{L}{F}-\LLie{L}{\rho(L_A)\OOp F}}\\
&&\!\!\!\! -\SScal{L}{\LLie{\aa(L\OOp F}{G}+\LLie{F}{\aa(L\OOp G}}\\
&\!\!\!\!=&\!\!\!\! \SScal{L\OOp\rho(L_A)}{\LLie{F}{G}}-\SScal{L}{\LLie{\rho(L_A)\OOp F+\aa(L\OOp F)}{G}+\\
&&\!\!\!\! \LLie{F}{\rho(L_A)\OOp G+\aa(L\OOp G)}}\,.
\EAS
In the previous calculation the results of Lemmas~\ref{lemmamio1} and \ref{lemmamio2} were used. One should remember that, being $\rho(L_A)$ a symmetric matrix, it holds for all $M\in\Gotico{gl}(n)$
$$
\rho(L_A)\OOp M = \rho(L_A)\Op M.
$$
The expression (\ref{Pois_Skl}) in Proposition~\ref{Skl_bra} is  thus proved.


\begin{thebibliography}{0}
\bibitem{tesiChiara} \textsc{Andr\`a C.}, \textit{Una struttura trihamiltoniana per il reticolo di 
Toda}, BSc thesis, University of Torino 2004.
%
\bibitem{CalaGonone2004} \textsc{Andr\`a C., Degiovanni L.}, \textit{New examples of trihamiltonian structures linking different Lenard chains}, in \textit{Symmetry and Perturbation Theory, proceedings of the international conference on SPT 2004}, World Scientific (2004), 13--21.
%
\bibitem{Bog1976} \textsc{Bogoyavlensky O.}, \textit{On perturbation tof the periodic Toda lattice},
Comm. Math. Phys., \textbf{51} (1976), 201--209.
%
\bibitem{Dam1994} \textsc{Damianou P.}, \textit{Multiple Hamiltonian structures for Toda-type systems},
J. Math. Phys., \textbf{35} (1994), 5511--5541.
%
\bibitem{DasO1989} \textsc{Das A., Okubo S.}, \textit{A systematic Study of the Toda Lattice},
Ann. Phys., \textbf{190} (1989), 215--232.
%
\bibitem{triham} \textsc{Degiovanni L., Magnano G.}, \textit{Tri-hamiltonian vector fields,
spectral curves and separation coordinates}, Rev. Math. Phys., \textbf{14} (2002), 1115--1163.
%
\bibitem{DLNT1986} \textsc{Deift P. A., Li L.-C., Nanda T., Tomei C.}, \textit{The Toda flow on a generic orbit is integrable},
Comm. Pure Appl. Math., \textbf{39} (1986), 183--232.
%
\bibitem{EFS1991} \textsc{Ercolani N. M., Flaschka H., Singer S.}, \textit{The geometry of the full Kostant-Toda lattice}, proceedings of \textit{Integrable systems (Luminy, 1991)}, Prog. Math. \textbf{115}, 181--225, Birkh\"auser 1993.
%
\bibitem{FMP2001} \textsc{Falqui G., Magri F., Pedroni M.}, \textit{Bihamiltonian geometry and 
separation of variables for Toda lattices}, J. Nonlin. Math. Phys., \textbf{8} (2001), 118--127.
%
\bibitem{BiHamSep} \textsc{Falqui G., Pedroni M.}, \textit{Separation of variables for
bi-hamiltonian systems}, Math. Phys. Anal. Geom., \textbf{6} (2003), 139--179.
%
\bibitem{Fla1974a} \textsc{Flaschka H.}, \textit{The Toda lattice. I. Existence of integrals},
Phys. Rev. B, \textbf{9} (1974), 1924--1925.
%
\bibitem{Hen1974} \textsc{H\'enon M.}, \textit{Integrals of the Toda lattice},
Phys. Rev. B, \textbf{9} (1974), 1921--1923.
%
\bibitem{Kos1979b} \textsc{Kostant B.}, \textit{The solution to a generalized Toda lattice and representation theory},
Adv. in Math., \textbf{34} (1979), 195--338.
%
\bibitem{Kup1985} \textsc{Kupershmidt B. A.}, \textit{Discrete Lax equations and dif\-fe\-ren\-tial
difference calculus}, Ast\'erisque, \textbf{123} (1985).
%
\bibitem{Mag2000} \textsc{Ibort A., Magri F., Marmo G.}, \textit{Bihamiltonian structures and
St\"ackel separability}, J. Geom. Phys., \textbf{33} (2000), 210--228.
%
\bibitem{M1997b} \textsc{Magri F.}, \textit{Eight lectures on integrable systems}, Lecture Notes in
Physics, \textbf{495} Springer Verlag, 1997.
%
\bibitem{Man1974} \textsc{Manakov S. P.}, \textit{Complete integrability and stochastization of discrete dynamical systems},
Sov. Phys. JETP, \textbf{40} (1974), 269--274.
%
\bibitem{McLS2000} \textsc{McLenaghan R., Smirnov R. G.}, \textit{Separabilty of the Toda lattice},
Appl. Math. Lett., \textbf{13} (2000), 77--82.
%
\bibitem{Meu2000} \textsc{Meucci A.}, \textit{Compatible Lie Algebroids and the Periodic Toda
Lattice}, J. Geom. Phys., \textbf{35} (2000), 273--287.
%
\bibitem{Meu2001} \textsc{Meucci A.}. \textit{Toda Equations, bi-Hamiltonian Systems, and Com\-pa\-ti\-ble
Lie Algebroids}, Math. Phys. Anal. Geom., \textbf{4} (2001), 131--146.
%
\bibitem{MorPiz1996a} \textsc{Morosi C., Pizzocchero L.}, \textit{$R$-matrix Theory, Formal Casimirs and the
Periodic Toda Lattice}, J. Math. Phys., \textbf{37} (1996), 4484--4513.
%
\bibitem{Mos1975b} \textsc{Moser J.}, \textit{Finitely many mass points on the line under the influence of an exponential potential --- An integrable system},
in \textit{Dynamical systems, theory and applications}, Lect. Notes in Phys. \textbf{38}, 467--497, Springer 1975.
%
\bibitem{ORagn1989}\textsc{Oevel W., Ragnisco O.}, \textit{$R$-matrices 
and Higher Poisson Brackets for Integrable Systems}, Phys. A, 
\textbf{161} (1989), 181--220.
%
\bibitem{ReySem1994} \textsc{Reyman A. G., Sem\"enov-Tian-Shansky M. A.}, \textit{Group theoretical
methods in the theory of finite dimensional integrable systems}, in \textit{Dynamical systems VII},
Enc. Math. Sci., \textbf{16}, 116--225, Springer-Verlag, 1994.
%
\bibitem{Rui1990} \textsc{Ruijsenaars S. N. M.}, \textit{Relativistic Toda systems},
Comm. Math. Phys., \textbf{133} (1990), 217--247.
%
\bibitem{Sem1982} \textsc{Sem\"enov-Tian-Shansky M. A.}, \textit{What is a classical $r$-matrix?},
Funct. Anal. Appl., \textbf{16} (1982), 263--270.
%
\end{thebibliography}
\end{document}